\documentclass[graybox]{svmult}

\usepackage{amssymb}
\usepackage{latexsym}
\usepackage{amsmath}

\usepackage{mathptmx}       
\usepackage{helvet}         
\usepackage{courier}        
\usepackage{type1cm}        
\usepackage{makeidx}         
\usepackage{graphicx}        
\usepackage{multicol}        
\usepackage[bottom]{footmisc}

\newcommand{\be}{\begin{eqnarray}}
\newcommand{\ee}{\end{eqnarray}}

\newcommand{\ket}[1]{\mbox{$\mid #1\,\rangle$}}

\newcommand{\pro}[2]{\mbox{$\langle\, #1 \mid #2\,\rangle$}}
\newcommand{\expec}[1]{\mbox{$\langle\, #1\,\rangle$}}

\renewcommand{\d}{\mbox{${\rm d}$}} 
\newcommand{\lp}{\ell_{\rm p}}
\newcommand{\mpl}{m_{\rm p}}
\newcommand{\gn}{G_{\rm N}}
\newcommand{\rh}{R_{\rm H}}

\makeindex             

\begin{document}

\title*{What is the Schwarzschild radius of a quantum mechanical particle?}
\author{Roberto Casadio}
\institute{Roberto Casadio \at Dipartimento di Fisica e Astronomia, Universit\`a di
Bologna, and I.N.F.N., Sezione di Bologna,
via Irnerio~46, 40126~Bologna, Italy, \email{casadio@bo.infn.it}
}
%
%
\maketitle
\abstract{A localised particle in Quantum Mechanics is described by a wave packet
in position space, regardless of its energy.
However, from the point of view of General Relativity, if the particle's energy density
exceeds a certain threshold, it should be a black hole.
In order to combine these two pictures, we introduce a horizon wave-function
determined by the position wave-function, which yields the probability that the particle
is a black hole.
The existence of a (fuzzy) minimum mass for black holes naturally follows, and we also show
that our construction entails an effective Generalised Uncertainty Principle
simply obtained by adding the uncertainties coming from the two wave-functions.}
\section{The Schwarzschild link}
In natural units, with $c=1$ (and $\hbar=\lp\,\mpl$), the Newton constant is given by
\be
\gn=\lp/\mpl
\ ,
\ee
where $\lp$ and $\mpl$ are the Planck length and mass, respectively,
and converts mass (or energy) into length.
This naive observation stands behind Thorne's {\em hoop conjecture\/}~\cite{Thorne:1972ji}:
A black hole forms when the impact parameter $b$ of two colliding objects
is shorter than the {\em Schwarzschild gravitational radius\/} of the system, that is for
\be
\rh
\equiv
2\,\lp\,\frac{E}{\mpl}
\gtrsim
b
\ ,
\label{hoop}
\ee
where $E$ is total energy in the centre-of-mass frame. 
The emergence of the Schwarz\-schild radius is indeed easy to understand in a spherically
symmetric space-time, where the metric $g_{\mu\nu}$ can be written as
\be
\d s^2
=
g_{ij}\,\d x^i\,\d x^j
+
r^2(x^i)\left(\d\theta^2+\sin^2\theta\,\d\phi^2\right)
\ ,
\label{metric}
\ee
with $x^i=(x^1,x^2)$ coordinates on surfaces of constant angles $\theta$ and $\phi$.
The location of a trapping horizon, a sphere where the escape velocity equals
the speed of light, is then determined by
\be
0
=
g^{ij}\,\nabla_i r\,\nabla_j r
=
1-\frac{2\,M}{r}
\ ,
\label{th}
\ee
where $\nabla_i r$ is the covector perpendicular to surfaces of constant area
$\mathcal{A}=4\,\pi\,r^2$.
The active gravitational (or Misner-Sharp) mass $M$ represents
the total energy enclosed within a sphere of radius $r$, and, if we set $x^1=t$
and $x^2=r$, is explicitly given by
\be
M(t,r)=\frac{4\,\pi\,\lp}{3\,\mpl}\int_0^r \rho(t, \bar r)\,\bar r^2\,\d \bar r
\ ,
\label{M}
\ee
where $\rho=\rho(x^i)$ is the matter density.
It is usually very difficult to follow the dynamics of a given matter distribution
and find surfaces satisfying Eq.~\eqref{th},
but an horizon exists if there are values of $r$ such that 
$\rh=2\,M(t,r)>r$, which is a mathematical reformulation of
the hoop conjecture~\eqref{hoop}.
\section{Horizon wave-function}
The hoop conjecture was formulated having in mind black holes of astrophysical
size~\cite{payne}, for which a classical metric and horizon structure
are reasonably safe concepts.
However, for elementary particles quantum effects may not be neglected~\cite{acmo}.
Consider a spin-less point-like source of mass $m$, whose Schwarzschild radius
is given by $\rh$ in Eq.~\eqref{hoop} with $E=m$.
The Heisenberg principle introduces an uncertainty in its spatial localisation,
of the order of the Compton-de~Broglie length,
$\lambda_m \simeq \lp\,{\mpl}/{m}$.
Assuming quantum physics is a more refined description of reality
implies that $\rh$ only makes sense if it is larger than $\lambda_m$,
\be
\rh\gtrsim \lambda_m
\quad
\Rightarrow
\quad
m
\gtrsim
\mpl
\quad
({\rm or}\ M\gtrsim\lp)
\ .
\label{clM}
\ee
Note that this argument employs the flat space Compton length,
and it is likely that the particle's self-gravity will affect it.
However, we can still assume the condition~\eqref{clM} holds as an
order of magnitude estimate, hence black holes can only exist with mass (much)
larger than the Planck scale.
\par
We are thus facing a deeply conceptual challenge:
how can we describe systems containing both quantum mechanical particles
and classical horizons?
For this purpose, we shall define a horizon wave-function that can be
associated with any localised quantum mechanical particle~\cite{fuzzy},
and that will put on quantitative grounds the condition~\eqref{clM}
that distinguishes black holes from regular particles.
\par
The quantum mechanical state representing and object, which is both
{\em localised in space\/} and {\em at rest\/} in the chosen reference frame,
must be described by a wave-function $\psi_{\rm S}\in L^2(\mathbb{R}^3)$,
which can be decomposed into energy eigenstates,
\be
\ket{\psi_{\rm S}}
=
\sum_E\,C(E)\,\ket{\psi_E}
\ ,
\ee
where the sum represents the spectral decomposition in Hamiltonian eigenmodes,
\be
\hat H\,\ket{\psi_E}=E\,\ket{\psi_E}
\ ,
\ee
and $H$ can be specified depending on the model we wish to consider.
If we also assume the state is {\em spherically symmetric\/}, we can
introduce a Schwarzschild radius $\rh=\rh(E)$ associated to each component
$\psi_E$ of energy $E$, by inverting Eq.~\eqref{hoop},
and define the (unnormalised) {\em horizon wave-function\/} as
\be
\tilde\psi_{\rm H}(\rh)
=
C\left(E
=
\mpl\,\frac{\rh}{2\,\lp}\right)
\ .
\ee
The normalisation is finally fixed by employing the inner product
\be
\pro{\psi_{\rm H}}{\phi_{\rm H}}
=
4\,\pi\,\int_0^\infty
\psi_{\rm H}^*(\rh)\,\phi_{\rm H}(\rh)\,\rh^2\,\d \rh
\ .
\ee
\par
We interpret the normalised wave-function $\psi_{\rm H}$ as yielding
the probability that we would detect a horizon of areal radius $r=\rh$ associated
with the particle in the quantum state $\psi_{\rm S}$.
Such a horizon is necessarily ``fuzzy'', like the particle's position,
unless the width of $\psi_{\rm H}$ is negligibly small.
Moreover, the probability density that the particle lies inside its own horizon of radius
$r=\rh$ will be given by
\be
P_<(r<\rh)
=
P_{\rm S}(r<\rh)\,P_{\rm H}(\rh)
\ ,
\label{PrlessH}
\ee
where
$
P_{\rm S}(r<\rh)
=
4\,\pi\,\int\limits_0^{\rh}
|\psi_{\rm S}(r)|^2\,r^2\,\d r
$
is the probability that the particle is inside the sphere of radius $r=\rh$,
and
$
P_{\rm H}(\rh)
=
4\,\pi\,\rh^2\,|\psi_{\rm H}(\rh)|^2
$
is the probability that the horizon is located on the sphere of radius $r=\rh$.
Finally, by integrating~\eqref{PrlessH} over all possible values of the radius,
\be
P_{\rm BH}
=
\int_0^\infty P_<(r<\rh)\,\d \rh
\ ,
\label{PBH}
\ee
the probability that the particle is a black hole will be obtained.
\subsection{Gaussian particle}
\begin{figure}[t]
\sidecaption
\includegraphics[width=7cm,height=4cm]{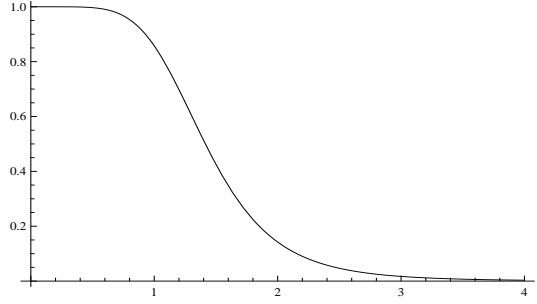}
%
\caption{Probability that particle of width $\ell$ is a black hole
as a function of $\ell/\lp$.
\label{prob}}
\end{figure}

The above construction can be straightforwardly applied to a particle
described by the Gaussian wave-function
\be
\psi_{\rm S}(r)
=
\frac{e^{-\frac{r^2}{2\,\ell^2}}}{\ell^{3/2}\,\pi^{3/4}}
\ ,
\ee
where the width $\ell\sim\lambda_{m}$.
This wave-function in position space corresponds to the momentum space wave-function
\be
\psi_{\rm S}(p)
=
\frac{e^{-\frac{p^2}{2\,\Delta^2}}}{\Delta^{3/2}\,\pi^{3/4}}\,
\ ,
\ee
where $p^2=\vec p\cdot\vec p$ and $\Delta=\hbar/\ell=\mpl\,\lp/\ell$.
For the energy of the particle, we simply assume the relativistic mass-shell relation
in flat space, $E^2=p^2+m^2$, and we easily obtain the normalised horizon wave-function
\be
\psi_H(\rh)
=
\frac{\ell^{3/2}\,e^{-\frac{\ell^2\,\rh^2}{8\,\lp^4}}}
{2^{3/2}\,\pi^{3/4}\,\lp^3}
\ .
\ee
Note that, since 
$\expec{\hat r^2}\simeq \ell^2$ and $\expec{\hat R_{\rm H}^2}\simeq \lp^4/\ell^2$,
we expect the particle will be inside its own horizon if $\expec{\hat r^2}\ll \expec{\hat R_{\rm H}^2}$,
which precisely yields the condition~\eqref{clM} if $\ell\simeq \lambda_m$.
In fact, the probability density~\eqref{PrlessH} can now be explicitly computed,
\be
P_<(r<\rh)
=
\frac{\ell^3\,\rh^2}{2\,\sqrt{\pi}\,\lp^6}\,
e^{-\frac{\ell^2\,\rh^2}{4\,\lp^4}}
\left[
{\rm Erf}\left(\frac{\rh}{\ell}\right)
-
\frac{2\,\rh}{\sqrt{\pi}\,\ell}\,
e^{-\frac{\rh^2}{\ell^2}}
\right]
\ ,
\label{Pin}
\ee
from which we derive the probability~\eqref{PBH} for the particle to be a black hole,
\be
P_{\rm BH}(\ell)
=
\frac{2}{\pi}\left[
\arctan\left(2\,\frac{\lp^2}{\ell^2}\right)
+
2\,\frac{\ell^2\,(4-\ell^4/\lp^4)}{\lp^2\,(4+\ell^4/\lp^4)^2}
\right]
\ .
\label{Pbh}
\ee
In Fig.~\ref{prob}, we show the probability~\eqref{Pbh} that the particle is a black hole as a function
of the Gaussian width $\ell$ (in units of $\lp$).
From the plot of $P_{\rm BH}$, it appears that the particle is most likely a black hole,
$P_{\rm BH}\simeq 1$, if $\ell\lesssim\lp$.
Assuming $\ell=\lambda_m=\lp\,\mpl/m$, we have thus derived a result in qualitative 
agreement with the condition~\eqref{clM}, but from a totally quantum mechanical picture. 
Strictly speaking, there is no black hole minimum mass in our treatment, but a vanishing probability
for a particle of ``small'' mass (say $m\lesssim \mpl/4$, that is $\ell\gtrsim 4\,\lp$), to be a black hole.   
\subsection{Generalised uncertainty principle}
\begin{figure}
\sidecaption
\includegraphics[width=7cm,height=4cm]{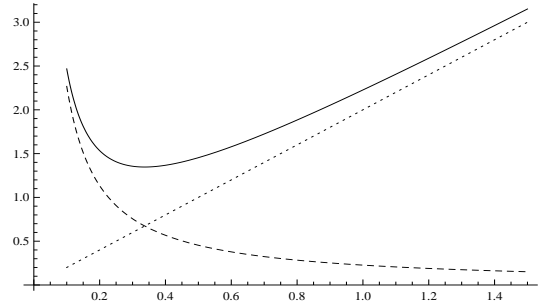}
%
\caption{Uncertainty relation~\eqref{effGUP} (solid line) as a combination
of the Quantum Mechanical uncertainty (dashed line) and the uncertainty
in horizon radius (dotted line) (lengths in units of $\lp$ and momentum in
units of $\mpl$).
\label{pGUP}}
\end{figure}

For the Gaussian packet described above, the Heisenberg
uncertainty in radial position is given by
\be
\expec{\Delta r^2}
=
4\,\pi\,\int_0^{\infty}
|\psi_{\rm S}(r)|^2\,r^4\,\d r
-
\left(
4\,\pi\,\int_0^{\infty}
|\psi_{\rm S}(r)|^2\,r^3\,\d r
\right)^2
=
\frac{3\,\pi-8}{2\,\pi}\,
\ell^2
\ ,
\label{Dr}
\ee
and, analogously, the uncertainty in the horizon radius will be given by
\be
\expec{\Delta \rh^2}
=
4\,\frac{3\,\pi-8}{2\,\pi}\,
\frac{\lp^4}{\ell^2}
\ .
\label{DRH}
\ee
Since
$
\expec{\Delta p^2}
=
\left(\frac{3\,\pi-8}{2\,\pi}\right)
\mpl^2\,\frac{\lp^2}{\ell^2}
\equiv
\Delta p^2
$,
we can also write
\be
\ell^2
=
\frac{3\,\pi-8}{2\,\pi}\,
\lp^2\,\frac{\mpl^2}{\Delta p^2}
\ .
\ee
Finally, by combining the uncertainty~\eqref{Dr} with \eqref{DRH} linearly,
we find
\be
\Delta r
\equiv
\sqrt{\expec{\Delta r^2}}
+
\gamma\,
\sqrt{\expec{\Delta \rh^2}}
=
\frac{3\,\pi-8}{2\,\pi}\,
\lp\,\frac{\mpl}{\Delta p}
+
2\,\gamma\,\lp\,\frac{\Delta p}{\mpl}
\ ,
\label{effGUP}
\ee
where $\gamma$ is a coefficient of order one, and the result is
plotted in Fig.~\ref{pGUP} (for $\gamma=1$).
This is precisely the kind of result one obtains from the generalised uncertainty principles
considered in Refs.~\cite{scardigli}, leading to a minimum measurable length
\be
\Delta r
\ge
2\,\sqrt{\gamma\,\frac{3\,\pi-8}{\pi}}\,\lp
\simeq
1.3\,\sqrt{\gamma}\,\lp
\ .
\ee
\par
Of course, one might consider different ways of combining the two
uncertainties~\eqref{Dr} and \eqref{DRH}, or even avoid this step and
just make a direct use of the horizon wave-function.
In this respect, the present approach appears more flexible and 
does {\em not\/} require modified commutators for the canonical
variables $r$ and $p$.
\section{Final remarks}
So far, the idea of the horizon wave-function was just applied to the very simple
case of a spinless massive particle, and expected results (existence of a minimum
black hole mass and generalised uncertainty relation) were recovered and refined~\cite{fuzzy}.
Next, it should be applied to more realistic systems.
For example, one could investigate dispersion relations derived from quantum field theory
in curved space-time, and a better definition of what a localised state in the latter context
should probably be employed as well~\cite{NW49}.
Regardless of such improvements, the conceptual usefulness of our construction should
already be clear, in that it allows us to deal with very quantum mechanical sources,
and to do so in a quantitative fashion.
For example, one could review the issue of quantum black holes~\cite{qbh} in light of
the above formalism, as well as finally tackle the description of black hole formation and
dynamical horizons in the gravitational collapse of truly quantum matter~\cite{acmo,qgc}.
\end{document}